\documentclass[11pt,a4paper]{article}

\usepackage[T1]{fontenc}
\usepackage[utf8]{inputenc}
\usepackage{newtxtext,newtxmath}
\usepackage{amsmath,bm}

\usepackage{graphicx}
\usepackage{geometry}
\usepackage{microtype}
\usepackage[numbers,sort&compress]{natbib}
\usepackage[font=small,labelfont=bf]{caption}
\usepackage[colorlinks=true,linkcolor=black,citecolor=blue,urlcolor=blue]{hyperref}

\hypersetup{
  pdftitle={Machine Learning Hamiltonians with Polarizations for Heterogeneous Systems},
  pdfauthor={Jia-Yu Zhou, Wentao Zhang, Mouyi Weng, Zhao-Yi Yan, Lin-Wang Wang}
}

\newcommand{\Eext}{\mathbf{E}_{\mathrm{ext}}}
\newcommand{\HSR}{H_{\mathrm{SR}}}
\newcommand{\HLR}{H_{\mathrm{LR}}}
\newcommand{\GI}{\mathrm{GI}}

\title{Machine Learning Hamiltonians with Polarizations for Heterogeneous Systems}
\author{%
Jia-Yu Zhou$^{1,2,*}$,\quad Wentao Zhang$^{1}$,\quad Mouyi Weng$^{1}$,\quad Zhao-Yi Yan$^{1}$,\quad Lin-Wang Wang$^{1,\dagger}$\\[0.85em]
\small $^{1}$\textit{State Key Laboratory of Optoelectronic Materials and Devices,}\\
\small \textit{Institute of Semiconductors, Chinese Academy of Sciences, Beijing, 100083, China}\\[0.3em]
\small $^{2}$\textit{University of Chinese Academy of Sciences, Beijing, 101408, China}
}
\date{}

\begin{document}
\maketitle
\footnotetext[1]{\href{mailto:jyzhou@semi.ac.cn}{jyzhou@semi.ac.cn}}
\footnotetext[2]{\href{mailto:lwwang@semi.ac.cn}{lwwang@semi.ac.cn}}

\begin{abstract}
Machine learning Hamiltonians use local atomic structures to determine local Hamiltonian matrix elements without carrying out self-consistent calculations. They have been used to study large systems and to carry out high-throughput screening for their electronic and optical properties. The prediction of the Hamiltonian is essentially the prediction of the electron electrostatic potential energy; however, for heterogeneous systems, this potential energy is often determined by faraway features, such as interface or surface dipoles, or even an applied voltage in a nanodevice. Thus, it cannot be determined by local atomic structures alone. In this sense, one needs to solve the electron electrostatic potential energy self-consistently using intrinsic atomic dipoles, the applied external potential energy, and polarizabilities. Fortunately, the intrinsic dipoles and polarizabilities are local properties, whereas the applied potential energy is a prescribed input. In this work, we propose DeepH-Polar, a physically informed machine learning Hamiltonian model that explicitly predicts intrinsic atomic dipoles and polarizability tensors and uses them to solve the electron electrostatic potential energy self-consistently. Benchmarks on polar GaN systems and crystalline-Si/amorphous-SiO$_2$ interfaces demonstrate improved Hamiltonian accuracy and reproduction of thickness-dependent electrostatic variations, as well as spatially nonuniform field-induced screening that models without polarization fail to capture. Our model also allows us to study nanoscale electronic devices. By iteratively solving the Hamiltonian eigenproblem and self-consistently updating free-carrier occupations and the electrostatic potential, we simulate an 8,194-atom gated semiconductor device.
\end{abstract}

\section{Introduction}

Machine learning Hamiltonians (MLHs) have been developed in recent years and have been used in various types of applications, including large-scale electronic-structure calculations, high-throughput materials screening, and predictions of optical properties~\citep{gong2023deeph,zhong2024universal,zhong2026unihamgnn,schwade2026hamster}. The essential point of a machine learning Hamiltonian is to use atomic-structure-based neural networks, such as graph neural networks (GNNs), to predict density functional theory (DFT) Hamiltonian matrix elements in a local atomic-orbital basis. In doing so, one can avoid repeated self-consistent DFT calculations and thus significantly improve efficiency, especially for large systems. The GNN models can be equivariant under the three-dimensional Euclidean group $E(3)$, ensuring the proper transformation of orbital Hamiltonian matrix elements under spatial rotations and inversion. Representative models, such as DeepH-E3, HamGNN, and DeePTB-E3, have enabled accurate and efficient predictions of electronic Hamiltonians across increasingly diverse atomic environments~\citep{gong2023deeph,zhong2023hamgnn,zhouyin2025slem,unke2021phisnet,yu2023qhnet,qian2026maceh}. However, most current MLH models are used for relatively homogeneous systems, such as moir\'{e}-pattern structures of two-dimensional materials~\citep{gong2023deeph}, periodic crystals~\citep{zhong2023hamgnn,zhong2024universal}, or small-cell amorphous systems~\citep{dai2026amorphous}, where spatially varying long-range electrostatic effects are often less pronounced. Predicting the Hamiltonian is essentially predicting its electron electrostatic potential energy $\phi(\mathbf r)$ once the local contributions have been determined. In a heterogeneous structure, or a structure under an external electric field, $\phi(\mathbf r)$ is a global quantity that is not solely determined by local atomic structures; for example, its value at one point can be affected by a distant surface dipole. In such systems, current MLH models will fail. To use MLH approaches for many interesting problems, such as nanoscale devices, one must account for this global dependence of $\phi(\mathbf r)$.

A recent attempt to address this problem is HamGNN-LR, which supplements an equivariant short-range Hamiltonian with a physically derived long-range Coulomb contribution~\citep{zhong2026hamgnnlr}. By mapping the electronic density matrix to effective atomic charges and evaluating the potential energy through reciprocal-space Ewald summation, HamGNN-LR can capture built-in electrostatic variations in polar slabs and heterostructures that are difficult to reconstruct using finite-range message passing alone. However, in this scheme, the effective atomic charges are locally determined quantities that lack an explicit response mechanism to long-range electric fields, for example when the thickness of an asymmetric slab with a surface dipole moment changes or when an electric field is applied in a device simulation. Without an explicit description of how these charges respond to the long-range electric field, a charge-based long-range correction alone will still have difficulty determining the screened Hamiltonian under different electrostatic conditions. Describing both built-in long-range electrostatics and electric-field-induced screening in MLH therefore remains an unsolved problem.

A similar approach to MLH is the charge-patching method (CPM)~\citep{wang2002patching}, which obtains atomic charge motifs from small-system DFT calculations that share the same local atomic environments as the large target system. Although the charge density is a more local property than the electron electrostatic potential energy $\phi(\mathbf r)$, it still faces the issue of dependence on global electric fields. This is similar to the situation in HamGNN-LR. Under a long-range electric field, the charge motif can be slightly different due to polarization, which must be taken into account. A subsequent solution is to include an atomic polarization motif, which, simply speaking, describes the change in the atomic charge motif under an additional electric field~\citep{cartoixa2005micro,wang2007motif}. CPM with polarization motifs can be used to study electronic devices and large-scale heterostructures, as demonstrated by the atomic motif-based Poisson solver (AMPS)~\citep{yan2025amps}. These studies demonstrate that including the intrinsic atomic charge distribution and its polarization response simultaneously is a reasonable and effective way to address this problem.

For materials with an electron energy gap, a natural way to include the polarization is to introduce atomic dipoles and their polarization response. Classical polarizable models, such as the Atomic Multipole Optimized Energetics for Biomolecular Applications (AMOEBA) force field, combine permanent atomic multipoles with self-consistently induced dipoles, thereby describing both intrinsic electrostatic interactions and polarization responses to electric fields~\citep{ponder2010amoeba}. Thole-type damping schemes are commonly used to regularize dipolar interactions at short range~\citep{thole1981dipole}. Similar ideas have also been incorporated into self-consistent machine-learning potentials~\citep{gao2022scfnn}. These studies illustrate how local atomic environments can characterize intrinsic electrostatic distributions and polarization-response parameters. A self-consistent solution for the polarization of atomic dipole moments by the long-range electric field, and the modification of this field by the dipoles in turn, can thus provide the electron electrostatic potential energy $\phi(\mathbf r)$. Incorporating this electron electrostatic potential energy into the Hamiltonian matrix elements can then provide the needed long-range correction.

Here we introduce DeepH-Polar, a polarization-aware extension of DeepH-E3 that incorporates environment-dependent intrinsic atomic dipoles $\mathbf p_i^0$ and polarizability tensors $\boldsymbol\alpha_i$. The formalism is based on a physical polarization model and is thus a physics-informed neural network (PINN) model~\citep{raissi2019pinn}. Its self-consistent long-range coupling through Poisson electrostatics provides a unified description of built-in electrostatic effects and atomically resolved screening under long-range electric fields. Analytical derivatives of the resulting Hamiltonian with respect to the applied field enable direct training using data from finite-field DFT calculations. We demonstrate that DeepH-Polar accurately captures both long-range electrostatic effects and atomically resolved field-induced screening across polar GaN bulk, GaN/AlN superlattices, GaN slabs, and crystalline-Si/amorphous-SiO$_2$ (c-Si/a-SiO$_2$) interfaces, including thickness-dependent electrostatic variations, spatially nonuniform field responses, and field-dependent electronic structures. Even for relatively small systems without obvious long-range electric fields, DeepH-Polar has smaller prediction errors than its DeepH-E3 counterpart.

We have also applied DeepH-Polar to electronic device simulations. Recently developed models such as DeePTB--NEGF and HamGNN--NEGF have demonstrated the feasibility of machine learning Hamiltonians for large-scale quantum-transport simulations~\citep{zou2025deeptbnegf,liu2026hamgnnnegf}. However, a unified atomistic treatment of long-range electrostatics and polarization screening has not been explicitly incorporated into these models. To illustrate the capability of DeepH-Polar, we calculate the relationship between carrier concentration and gate voltage as well as how the channel band-edge states evolve with gate voltage for an 8,194-atom nanodevice, using our polarization model in the Poisson equation and the predicted Hamiltonian to obtain electronic eigenstates.

\section{Results}
\subsection{DeepH-Polar framework and Hamiltonian prediction accuracy}

\begin{figure}[!htbp]
\centering
\includegraphics[width=\linewidth,height=0.56\textheight,keepaspectratio]{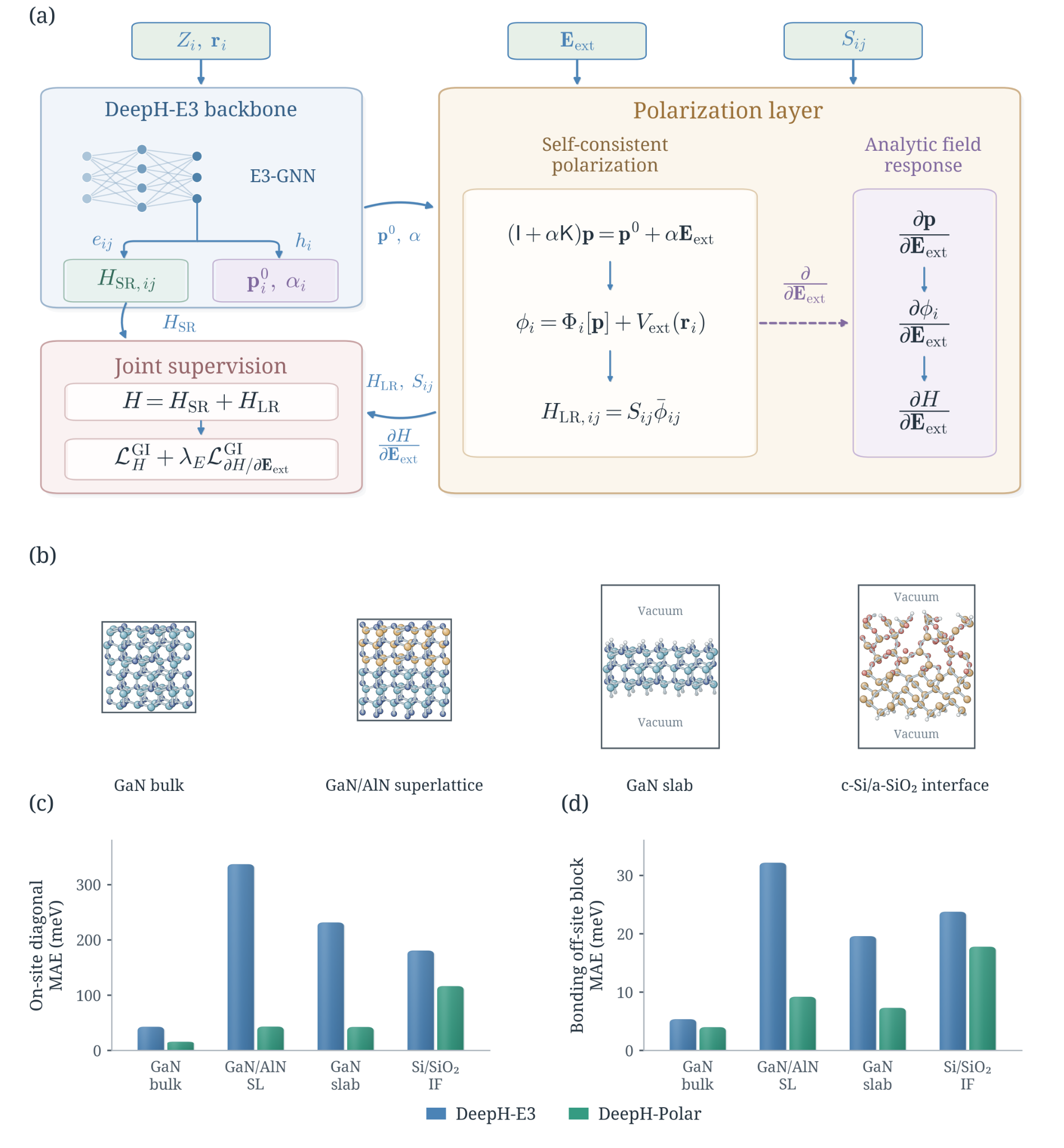}
\caption{\textbf{DeepH-Polar framework and Hamiltonian prediction accuracy.} \textbf{(a)} Schematic architecture of DeepH-Polar. The DeepH-E3 backbone predicts the short-range Hamiltonian, together with environment-dependent intrinsic atomic dipoles and polarizabilities. The polarization layer self-consistently determines the atomic dipoles and the resulting electron electrostatic potential energies, which are incorporated into the Hamiltonian as long-range contributions through the overlap matrix. Analytically evaluated field-response derivatives enable joint supervision of the Hamiltonian and its electric-field response when external-field training data are available. \textbf{(b)} Representative atomic structures of the four benchmark systems: GaN bulk, GaN/AlN superlattice, GaN slab, and c-Si/a-SiO$_2$ heterostructure. \textbf{(c,d)} Comparison of the MAEs of DeepH-E3 and DeepH-Polar for (c) on-site diagonal Hamiltonian matrix elements and (d) nearest-neighbor bonding off-site Hamiltonian blocks. All errors are evaluated on zero-field test structures after gauge alignment with the corresponding DFT references.}
\label{fig:framework}
\end{figure}

MLHs based on local atomic environments have a finite spatial cutoff and therefore cannot fully capture long-range electrostatic interactions and the screening by electronic charge. To address these limitations, we introduce \textbf{DeepH-Polar}, a polarization-aware extension of the gauge-invariant DeepH-E3 framework~\citep{gong2023deeph}. As illustrated in Fig.~\ref{fig:framework}a, using a neural network based on the DeepH-E3 backbone, DeepH-Polar predicts the short-range Hamiltonian $\HSR$, together with environment-dependent intrinsic atomic dipoles $\mathbf p_i^0$ and polarizability tensors $\boldsymbol\alpha_i$.

Rather than using effective atomic monopole charges to represent long-range electrostatics, as in charge-based schemes such as HamGNN-LR~\citep{zhong2026hamgnnlr}, we employ atomic dipoles. Depending on the charge-partitioning scheme, the atomic monopole-charge and dipole representations can be equivalent descriptions of the long-range electrostatic effect. However, a monopole-charge representation can involve long-range charge transfer, which is difficult to describe solely from local atomic structures. By assigning zero net monopole charge to each atomic polarization source, the dipolar representation avoids such explicit charge transfer and the need to impose a global charge-conservation constraint on the learned polarization sources. For example, with adjustable atomic partition parameters, a Hirshfeld-type scheme~\citep{hirshfeld1977fragments} can be used to assign zero net charge and a uniquely defined dipole moment to each atom. The dipole representation also provides a natural linear response to the local electric field. Here, each Gaussian polarization source has zero net charge by construction.

To make the field contributions clear, we separate the long-range electric field into the field $\mathbf E_i^{\mathrm{dip}}(\mathbf p)$ generated by the atomic dipoles and the additional, unscreened external field $\Eext$. Assuming that the atomic polarizability is represented by a site-dependent tensor $\boldsymbol\alpha_i$, the atomic dipole is
\begin{equation}
\label{eq:local-polarization}
\mathbf p_i=\mathbf p_i^0+\boldsymbol\alpha_i\left[\mathbf E_i^{\mathrm{dip}}(\mathbf p)+\Eext\right].
\end{equation}
Here, $i$ denotes the atomic index, and $\mathbf p_i^0$ is the intrinsic dipole in the absence of the modeled long-range electric field. Only the long-range dipole-induced field is explicitly included in $\mathbf E_i^{\mathrm{dip}}$; short-range electronic effects are represented by the learned local parameters and $\HSR$. The field from the dipoles is obtained by solving the linear Poisson equation for Gaussian-broadened polarization sources. Its linear action on the dipoles can be written as $\mathbf E_i^{\mathrm{dip}}(\mathbf p)=-\sum_j\mathbf K_{\sigma,ij}\mathbf p_j$, where $\mathbf K_\sigma$ is the Gaussian-regularized dipole-interaction kernel and $\sigma$ is the broadening width. The polarization equation therefore becomes
\begin{equation}
\label{eq:polarization}
\left(\mathbf I+\boldsymbol{\alpha}\mathbf K_{\sigma}\right)\mathbf p
=\mathbf p^0+\boldsymbol{\alpha}\Eext.
\end{equation}
The electron electrostatic potential energy at atom $i$ due to the long-range dipoles and the applied field is
\begin{equation}
\label{eq:potential}
\phi_i=\Phi_i[\mathbf p;\sigma]+V_{\mathrm{ext}}(\mathbf R_i).
\end{equation}
Here, $\Phi_i[\mathbf p;\sigma]$ is the \emph{electron electrostatic potential energy} induced by the dipoles, calculated from their Gaussian representation using Poisson electrostatics, and $V_{\mathrm{ext}}$ is the prescribed external electron potential energy corresponding to $\Eext$. In particular, $\phi_i$ represents the \emph{long-range correction}, not the full microscopic electron potential energy: local bonding and other short-range electronic contributions are already incorporated into $\HSR$. The complete Gaussian representation, Poisson equation, and periodic reciprocal-space expressions for $\mathbf K_\sigma$ and $\Phi_i$ are given in Methods, Eqs.~\eqref{eq:method-gaussian}--\eqref{eq:method-periodic-k}.

After $\phi_i$ is obtained, the long-range part of the Hamiltonian is
\begin{align}
\label{eq:hamiltonian}
H_{\mathrm{LR},ij}&=S_{ij}\bar\phi_{ij},\qquad
\bar\phi_{ij}=m_{Z_i Z_j}\phi_i+(1-m_{Z_i Z_j})\phi_j,\\
\label{eq:hamiltonian-total}
H&=\HSR+\HLR.
\end{align}
Here, $S_{ij}$ is the overlap-matrix block between the basis functions on atoms $i$ and $j$, and $m_{Z_iZ_j}$ is an element-pair interpolation coefficient for distributing the atomic potential energies over the corresponding Hamiltonian block. For on-site blocks $m_{AA}=1/2$. To preserve Hermiticity, the reverse ordering uses $m_{BA}=1-m_{AB}$. All training structures, including slabs containing vacuum, use periodic boundary conditions, so $\mathbf K_\sigma$ and $\Phi_i$ are evaluated in reciprocal space (Methods, Eqs.~\eqref{eq:method-periodic-phi} and \eqref{eq:method-periodic-k}). The common Gaussian broadening width $\sigma$ and the element-pair coefficients $m_{AB}$ are trainable. Together, these equations incorporate long-range electrostatic information through the Poisson solution while treating intrinsic and external-field-induced polarization within a single framework.

Analytically differentiating the polarization equations with respect to the external field gives the Hamiltonian response $\partial H/\partial\Eext$, which is evaluated explicitly and used for joint supervision:
\begin{equation}
\label{eq:loss}
\mathcal L=\mathcal L_H^{\GI}
+\lambda_E\mathcal L_{\partial H/\partial\Eext}^{\GI}.
\end{equation}
Here, $\lambda_E$ controls the relative contribution of field-response supervision and is set to zero when no external-field reference data are available. The superscript GI denotes gauge-invariant supervision: changing the uniform energy reference adds a constant times the overlap matrix $S$ to the Hamiltonian, and an optimal such shift is removed when evaluating the loss~\citep{wang2024universal}, ensuring consistent Hamiltonian comparisons for the polar systems and applied electric fields. The actual gauge-aligned loss, training weights, and analytic response equations are given in Methods, Eqs.~\eqref{eq:method-gi-loss}--\eqref{eq:method-field-label}. Through this joint supervision, the model directly learns the self-consistent induced-dipole contribution to electronic screening. We do not explicitly fit atomic dipole moments or long-range electric fields; rather, their effects are learned through the self-consistently calculated Hamiltonian matrix elements and, when available, its field response.

To assess the performance of DeepH-Polar across distinct chemical and electrostatic environments, we construct four benchmark datasets: GaN bulk, GaN/AlN superlattices, GaN slabs, and c-Si/a-SiO$_2$ heterostructures (Fig.~\ref{fig:framework}b), containing 598, 349, 251, and 600 structures, respectively. The structures were sampled from molecular-dynamics trajectories generated using a machine-learning force field (MLFF) and divided into training, validation, and test subsets in an approximately 10:1:1 ratio. Details of dataset generation, molecular-dynamics sampling, and DFT calculations are given in Methods. These datasets span polar bulk structures, interfaces, and surfaces, as well as chemically heterogeneous semiconductor--oxide interfaces, where long-range electrostatic fields can play an important role. DeepH-Polar consistently lowers the mean absolute error (MAE) for both on-site diagonal and nearest-neighbor bonding off-site Hamiltonian matrix elements relative to DeepH-E3 (Fig.~\ref{fig:framework}c,d). The improvement is especially pronounced for GaN/AlN superlattices and GaN slabs: their on-site diagonal MAEs decrease from 336.40 to 42.69\,meV and from 231.03 to 41.99\,meV, respectively. These gains are consistent with strong long-range electrostatic effects associated with polar interfaces and surfaces, which are examined more closely in the next section. All benchmarks presented here are evaluated at zero applied electric field, demonstrating the effectiveness of our framework in describing intrinsic long-range electrostatic effects, while its ability to capture external-field screening is examined separately below.

\subsection{Long-range electric field effect in polar superlattices and surfaces}

\begin{figure}[!htbp]
\centering
\includegraphics[width=\linewidth,height=0.56\textheight,keepaspectratio]{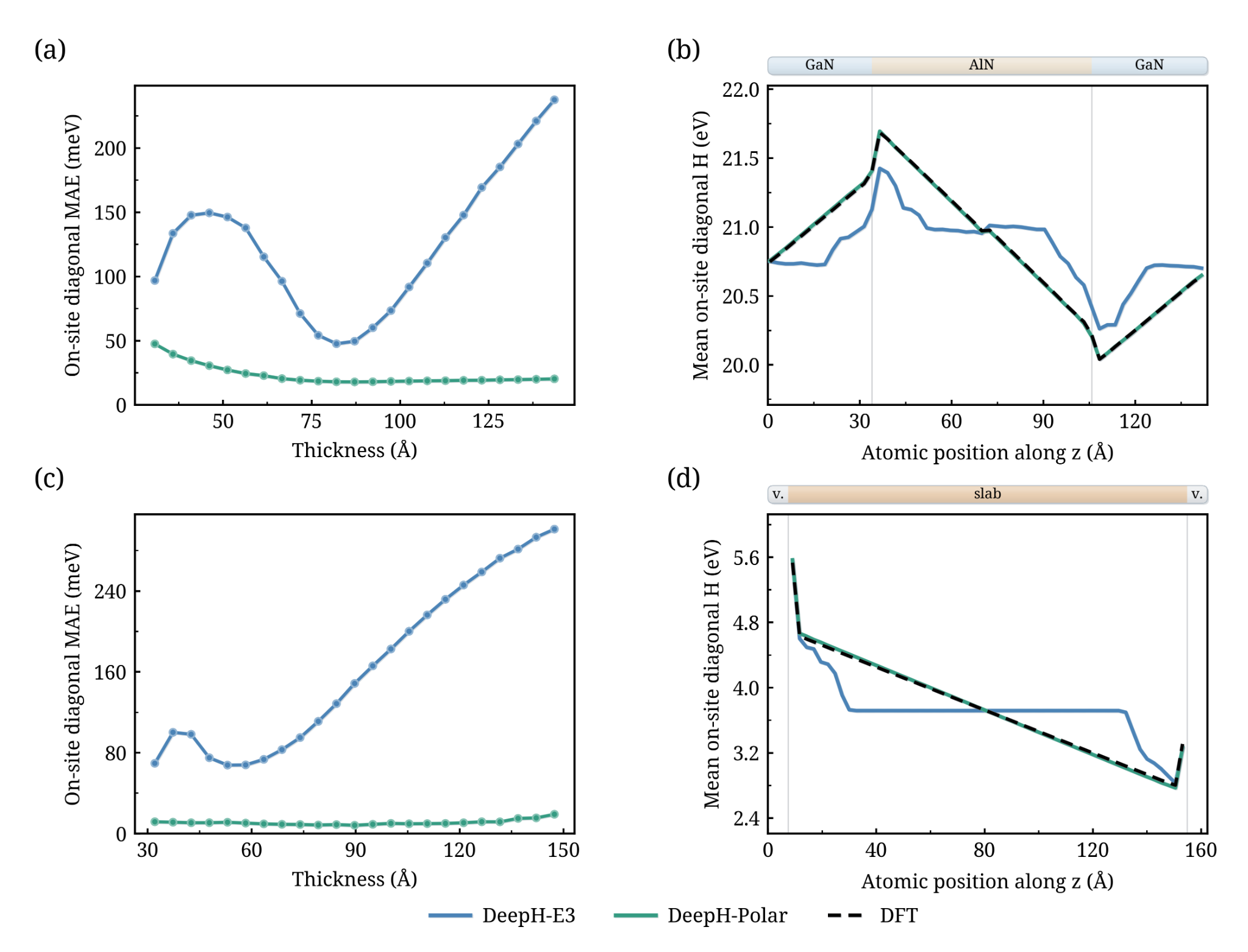}
\caption{\textbf{Long-range electrostatic effects in polar superlattices and surfaces.} \textbf{(a,c)} Thickness dependence of the on-site diagonal Hamiltonian MAEs for (a) GaN/AlN superlattices and (c) GaN slabs. \textbf{(b,d)} Spatial distributions of the mean on-site diagonal Hamiltonian matrix elements for 56-layer-thick structures of (b) GaN/AlN superlattice and (d) GaN slab, compared with DFT references. The values are averaged over N atoms in (b) and Ga atoms in (d). All Hamiltonian values are gauge-aligned with the corresponding DFT references. DeepH-Polar accurately reproduces the spatial variations in the on-site Hamiltonian dominated by long-range electrostatic effects and maintains consistently low errors across thicknesses, whereas DeepH-E3 errors increase sharply in thicker systems.}
\label{fig:longrange}
\end{figure}

To examine thickness-dependent long-range electric field effects, we consider ideal GaN/AlN superlattices and GaN slabs with increasing thickness along the polar axis (Fig.~\ref{fig:longrange}). Polarization discontinuities at the GaN/AlN interfaces and polar surface terminations produce internal electrostatic fields that extend well beyond the range of local atomic environments~\citep{bernardini1997spontaneous}. As the thickness increases, the associated electrostatic potential energy can vary over a larger distance, even though the local atomic coordination in the interior remains essentially unchanged. Consequently, a local Hamiltonian model cannot distinguish atoms whose nearby environments are the same but whose long-range electrostatic surroundings differ. In contrast, DeepH-Polar incorporates the potential energies generated by collective atomic dipoles through the Poisson solution, making the Hamiltonian sensitive to distant interface and surface dipoles.

As expected, the contrast between the two models becomes particularly pronounced for thicker structures (Fig.~\ref{fig:longrange}a,c). Although the DeepH-E3 on-site diagonal MAE is nonmonotonic at smaller thicknesses, it rises sharply at larger thicknesses, exceeding 200\,meV for the thickest GaN/AlN superlattices and approaching 300\,meV for the thickest GaN slabs. By comparison, DeepH-Polar keeps having low errors, approaching approximately 20\,meV in thicker superlattices and remaining around 10--20\,meV for most slab thicknesses. The origin of this difference is particularly clear in the 56-layer GaN slab (Fig.~\ref{fig:longrange}d). The DeepH-E3 backbone uses a cutoff radius of 7.2\,\text{\AA} and two hidden message-passing layers, yielding a maximum geometric receptive-field range of approximately 21.6\,\text{\AA} for the on-site representation. Deep inside a thick slab, Ga atoms therefore present essentially identical local environments to DeepH-E3, which predicts a constant on-site Hamiltonian across the interior. DFT, however, exhibits an approximately linear spatial variation arising from the long-range electrostatic potential, which is reproduced by DeepH-Polar through its self-consistent polarization and Poisson solution. The GaN/AlN superlattice exhibits the same limitation in a different geometry (Fig.~\ref{fig:longrange}b). These results indicate that electrostatic interactions dominate the long-range contributions to the Hamiltonian in polar systems and that DeepH-Polar captures these contributions accurately through its physics-based polarization model.

\subsection{\texorpdfstring{Electric-field response and screening in c-Si/a-SiO$_2$ heterostructures}{Electric-field response and screening in c-Si/a-SiO2 heterostructures}}

\begin{figure}[!htbp]
\centering
\includegraphics[width=\linewidth,height=0.56\textheight,keepaspectratio]{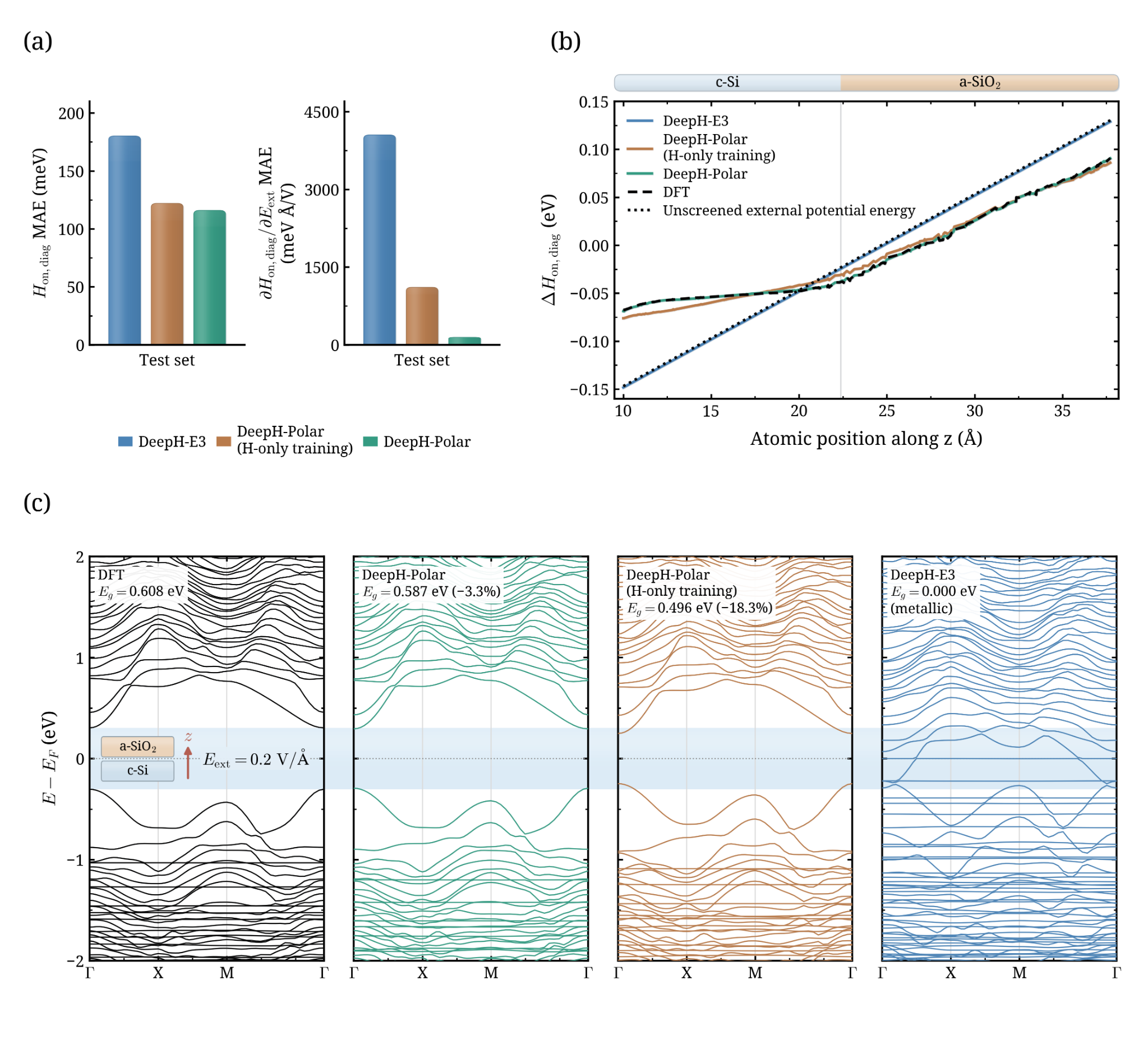}
\caption{\textbf{Electric-field response and screening in c-Si/a-SiO$_2$ heterostructures.} \textbf{(a)} Comparison of the MAEs in zero-field on-site diagonal Hamiltonian elements and their electric-field derivatives for DeepH-E3, DeepH-Polar trained without field-response supervision (H-only training), and the full DeepH-Polar model. \textbf{(b)} Spatial distribution of the field-induced on-site Hamiltonian response, $\Delta H_{\mathrm{on,diag}}=H_{\mathrm{on,diag}}(+E)-H_{\mathrm{on,diag}}(-E)$, with $E=0.005\,\mathrm{V}/\text{\AA}$, compared with DFT and the unscreened external potential energy. \textbf{(c)} Electronic band structures under a strong external field of $0.2\,\mathrm{V}/\text{\AA}$, showing band gaps for DFT, DeepH-Polar, and DeepH-Polar (H-only training), while DeepH-E3 incorrectly predicts metallic states. Hamiltonians are independently gauge-aligned with the corresponding DFT references at each electric-field strength. Incorporating electric-field response supervision substantially improves the accuracy of atomistic screening predictions, enabling DeepH-Polar to reproduce the field-dependent electronic structure even at field strengths beyond those used for training.}
\label{fig:field}
\end{figure}

Beyond its ability to reproduce intrinsic long-range electrostatics, DeepH-Polar can also describe electronic responses to externally applied electric fields. We demonstrate this capability using a device-relevant c-Si/a-SiO$_2$ heterostructure dataset (Fig.~\ref{fig:field}). Although the self-consistent polarization layer provides a physically motivated model of induced dipoles, zero-field Hamiltonian training alone may not have sufficient data to determine the parameters of the model. To isolate the effect of field-response supervision, we independently train a new DeepH-Polar model from scratch using only the zero-field Hamiltonian data. This control model, denoted DeepH-Polar (H-only training), uses the same zero-field training structures as the field-supervised model but excludes all finite-field response labels. The field-supervised model is included in the zero-field benchmarks of Fig.~\ref{fig:framework}c,d. Further details of the field pairs and the model training protocol are given in Methods. In the full model, the field derivative $\partial H/\partial\Eext$ is evaluated analytically from the polarization equations and supervised using central finite differences of DFT Hamiltonians under fields of $\pm0.005\,\mathrm{V}/\text{\AA}$ along the applied Cartesian directions, with each field pair providing one response label. As shown in Fig.~\ref{fig:field}a, the two DeepH-Polar models achieve comparable zero-field on-site diagonal MAEs of 122.05 and 115.97\,meV for H-only and field-supervised training, respectively. In contrast, the corresponding field-response MAE is reduced from 1111.39 to 143.94\,meV\,\text{\AA}/V, compared with 4053.00\,meV\,\text{\AA}/V for DeepH-E3 with a directly added external potential. Thus, training on zero-field Hamiltonians alone can recover part of the field response, whereas accurate prediction of atomistic screening requires explicit field-response supervision.

The spatial character of this improvement is illustrated using a c-Si/a-SiO$_2$ structure held out entirely from training (Fig.~\ref{fig:field}b). We compare the on-site diagonal Hamiltonian differences $\Delta H_{\mathrm{on,diag}}(z)=H_{\mathrm{on,diag}}(z,+E)-H_{\mathrm{on,diag}}(z,-E)$ at $E=0.005\,\mathrm{V}/\text{\AA}$. The DFT response differs substantially from the unscreened external-potential-energy difference, exhibiting a weaker spatial gradient in c-Si and a steeper one in a-SiO$_2$, consistent with spatially nonuniform screening across the interface. In construction, DeepH-E3 adds only the external potential energy and therefore gives the unscreened response. DeepH-Polar with H-only training recovers part of the screening but shows noticeable deviations, whereas the field-supervised DeepH-Polar closely reproduces the DFT response across both regions and the interface. Accurate screening is particularly important for electronic device simulations.

We further test whether a model trained using small-field reference data can accurately predict electronic structures under substantially stronger applied fields (Fig.~\ref{fig:field}c). At $E_{\mathrm{ext},z}=0.2\,\mathrm{V}/\text{\AA}$, DeepH-Polar closely reproduces the DFT band dispersion and predicts a band gap of 0.587\,eV, compared with the DFT value of 0.608\,eV. The H-only model predicts a smaller gap of 0.496\,eV, while DeepH-E3 incorrectly yields unphysical metallic states. The test field is 40 times stronger than the magnitudes used for response supervision, and the atomic structure is the same held-out structure as in Fig.~\ref{fig:field}b. This indicates that even at high electric field, the linear response polarization model remains an effective model. Together, these results highlight the importance of explicit field-response supervision for accurate atomistic screening. They also suggest promising generalization across atomic structures and external-field conditions, providing a foundation for modeling complex electrostatic environments in large-scale semiconductor devices.

\subsection{Self-consistent atomistic electronic structure of a large-scale semiconductor device}

\begin{figure}[!htbp]
\centering
\includegraphics[width=\linewidth,height=0.56\textheight,keepaspectratio]{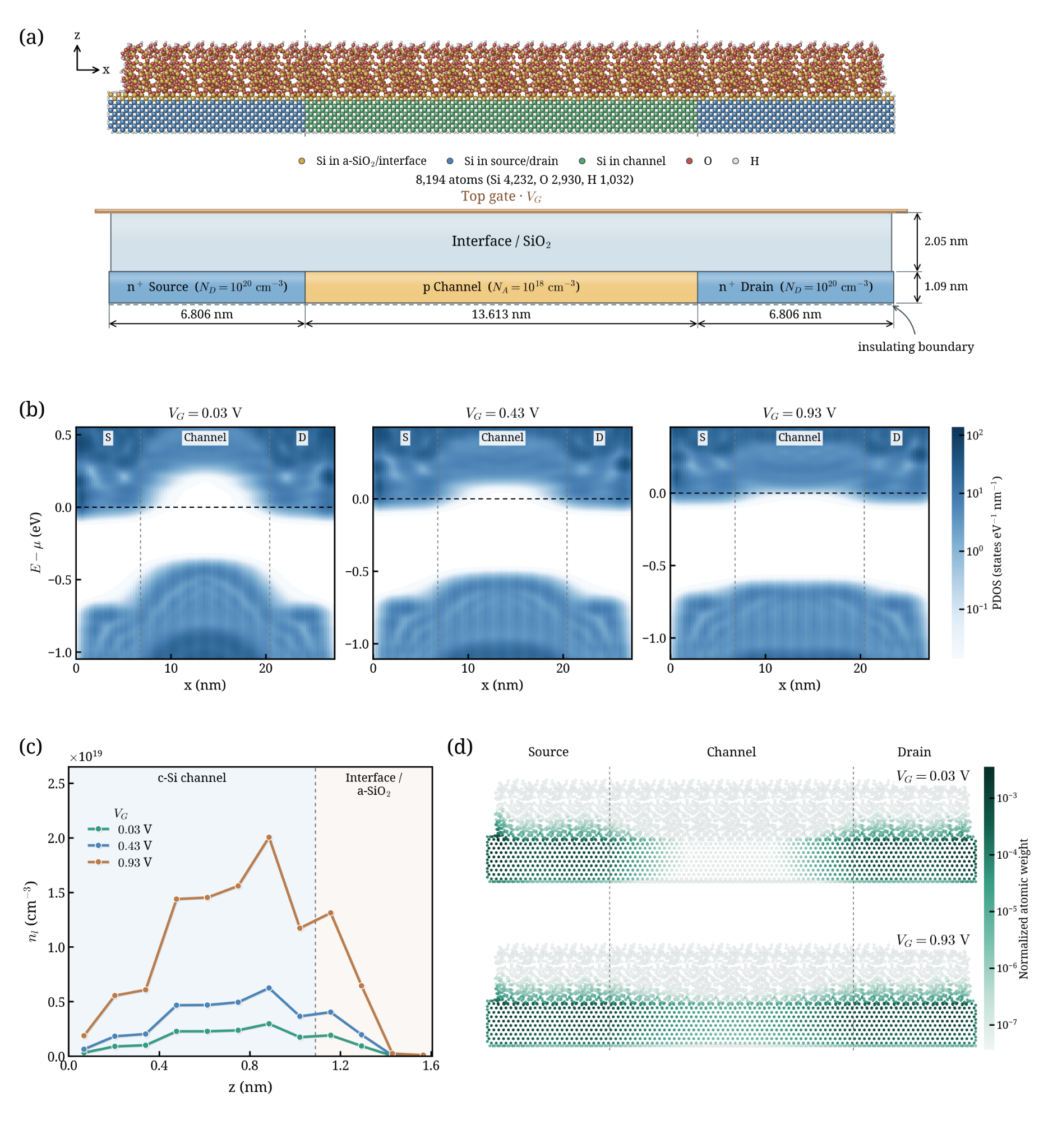}
\caption{\textbf{Self-consistent atomistic electronic structure of a large-scale semiconductor device.} \textbf{(a)} Atomistic structure and schematic geometry of the gate-controlled c-Si/a-SiO$_2$ device containing 8,194 atoms, with n-type source and drain regions and a p-type channel. \textbf{(b)} Gate-dependent spatially resolved PDOS along the source--channel--drain direction at $V_G=0.03$, $0.43$, and $0.93$\,V. \textbf{(c)} Atomic-layer-resolved free-electron densities across the c-Si channel and c-Si/a-SiO$_2$ interface at the corresponding gate voltages, revealing nonuniform carrier distributions and electronic penetration into the interfacial region. \textbf{(d)} Spatial distributions of atomic L\"owdin weights for four conduction-edge eigenstates within 50\,meV above the conduction-band minimum at $V_G=0.03$ and $0.93$\,V. The machine-learned Hamiltonian is diagonalized using FSM to obtain the electronic states and PDOS, from which the free-carrier density is calculated and fed back to self-consistently update the electrostatic potential and Hamiltonian at $V_{DS}=0$ and $T=300$\,K.}
\label{fig:device}
\end{figure}

To demonstrate the applicability of DeepH-Polar at device-relevant length scales, we perform self-consistent atomistic electronic-structure calculations for a gate-controlled c-Si/a-SiO$_2$ device comprising 8,194 atoms (4,232 Si, 2,930 O, and 1,032 H; Fig.~\ref{fig:device}a). The device contains an n-type source, a 13.613-nm-long p-type channel, and an n-type drain, with a 1.09-nm-thick c-Si layer beneath an approximately 2.05-nm-thick amorphous oxide. At $V_{DS}=0$ and $T=300\,$K, the self-consistent calculation starts from the atomic electron potential energies $\{\phi_i\}$, which are used to construct the long-range correction and total Hamiltonian using Eqs.~\eqref{eq:hamiltonian}--\eqref{eq:hamiltonian-total}. We transform the Hamiltonian into the L\"owdin-orthogonalized representation and obtain the relevant electronic eigenstates using the folded-spectrum method (FSM)~\citep{wang1994foldedspectrum} (Methods, Eq.~\eqref{eq:device-fsm}):
\begin{align}
\label{eq:device-diag}
\widetilde H &= S^{-1/2} H S^{-1/2}, &
\widetilde H\mathbf v_n &= \varepsilon_n\mathbf v_n,\\
\label{eq:device-proj}
w_{in} &= \sum_{a\in i}|v_{an}|^2, &
D_i(E) &= \sum_n w_{in}\delta_\eta(E-\varepsilon_n),
\end{align}
where $w_{in}$ is the atomic L\"owdin weight of eigenstate $n$ and $D_i(E)$ is the projected density of states (PDOS), with $\delta_\eta$ denoting the spectral broadening. The equilibrium chemical potential is fixed by the net-carrier constraint in the distant source--drain reservoirs (Methods, Eq.~\eqref{eq:device-mu}). Using the Fermi--Dirac occupations of the resulting eigenstates, we determine the atom-resolved electron populations $n_i^{\mathrm e}$, hole populations $n_i^{\mathrm h}$, and their net free-carrier charges $q_{i,\mathrm{free}}$ (Methods, Eq.~\eqref{eq:device-pop}). These atomic charges are Gaussian-broadened to construct the spatial free-charge density,
\begin{equation}
\label{eq:device-rho}
\rho_{\mathrm{free}}(\mathbf r)=\sum_i q_{i,\mathrm{free}}g_\sigma(\mathbf r-\mathbf R_i).
\end{equation}
The free-carrier and fixed-doping charge densities, $\rho_{\mathrm{free}}$ and $\rho_{\mathrm{fix}}$, determine the device electrostatic potential $\varphi_{\mathrm{ext}}$ through Poisson's equation subject to the gate and contact boundary conditions:
\begin{equation}
\label{eq:device-poisson}
-\varepsilon_0\nabla^2\varphi_{\mathrm{ext}}(\mathbf r)
=\rho_{\mathrm{free}}(\mathbf r)+\rho_{\mathrm{fix}}(\mathbf r),
\qquad V_{\mathrm{ext}}(\mathbf r)=-e\varphi_{\mathrm{ext}}(\mathbf r).
\end{equation}
With $\mathbf E_{\mathrm{ext}}(\mathbf R_i)=-\nabla\varphi_{\mathrm{ext}}(\mathbf R_i)$, the polarization and the next-iteration atomic electron potential energy are coupled by
\begin{equation}
\label{eq:device-polarization}
\begin{aligned}
(\mathbf I+\boldsymbol\alpha\mathbf K_{\sigma,\mathrm{device}})\mathbf p
&=\mathbf p^0+\boldsymbol\alpha\mathbf E_{\mathrm{ext}},\\
\phi_i&=\Phi_{i,\mathrm{device}}[\mathbf p;\sigma]+V_{\mathrm{ext}}(\mathbf R_i).
\end{aligned}
\end{equation}
For the device, we replace the periodic interaction kernel $\mathbf K_\sigma$ by $\mathbf K_{\sigma,\mathrm{device}}$, obtained from the electrostatic Green function with the device boundary conditions. The corresponding dipole-induced electron potential energy is $\Phi_{i,\mathrm{device}}[\mathbf p;\sigma]$, defined using the same boundary conditions. The two quantities are the device-specific forms of $\mathbf K_\sigma$ and $\Phi_i$ defined in Methods, Eqs.~\eqref{eq:method-green-k} and \eqref{eq:method-green-phi}.
The Hamiltonian--charge--polarization cycle is repeated until both $\phi_i$ and $q_{i,\mathrm{free}}$ satisfy the convergence thresholds. Details of the boundary conditions, charge-neutrality constraint, mixing strategy, and convergence tolerances are given in Methods.

The resulting self-consistent electronic structure reveals a gate-induced evolution of the channel states. As $V_G$ increases from $0.03$ to $0.93\,$V, the spatially resolved PDOS (Fig.~\ref{fig:device}b) shows a progressive downward shift of the channel conduction-band edge relative to the common chemical potential. The corresponding atomic-layer-resolved free-electron densities (Fig.~\ref{fig:device}c) increase with gate voltage and exhibit pronounced variations across the 1.09-nm channel thickness, with enhanced occupation toward the upper Si interface and finite electronic occupation extending into the c-Si/a-SiO$_2$ interfacial region. This penetration reflects an atomically resolved electronic distribution rather than a carrier density artificially terminated at a geometrical dielectric boundary. The eigenstate projections provide further evidence for the evolution toward a conducting channel: at $V_G=0.03\,$V, the selected conduction-edge states have little weight in the central channel, whereas at $V_G=0.93\,$V their atomic L\"owdin weights extend much more broadly across the channel (Fig.~\ref{fig:device}d). This redistribution is consistent with the progressive formation of an electronically accessible channel as the gate bias increases. These results highlight the potential of DeepH-Polar to bring first-principles-level electronic-structure accuracy, at manageable computational cost, to self-consistent simulations of nanoscale semiconductor devices approaching ten thousand atoms, thereby enabling atomistic resolution of quantum electronic effects.

\section{Discussion}
Electronic Hamiltonians in polar and interfacial systems depend on more than nearby atomic coordination: their spatial variation is governed by electrostatic fields originating throughout a structure and by the response of bound electrons to these fields. Built-in electrostatic effects and externally induced screening share the same underlying polarization physics, although the sources of the electrostatic fields differ. DeepH-Polar expresses this connection through environment-dependent intrinsic atomic dipoles and polarizabilities coupled by self-consistent Poisson electrostatics. The resulting framework treats long-range internal electrostatics and responses to externally applied electric fields within the same Hamiltonian formulation, rather than introducing separate corrections for the two effects. Compared with charge-based long-range Hamiltonian approaches such as HamGNN-LR~\citep{zhong2026hamgnnlr}, the dipolar formulation provides an explicit linear local-field response and preserves the overall neutrality of the polarization-induced charge distribution.

The benchmarks reveal two complementary aspects of this unified description. In polar GaN/AlN superlattices and GaN slabs, DeepH-Polar recovers the long-range spatial variation of the Hamiltonian and maintains low prediction errors as the thickness increases, even when atoms deep inside a structure have indistinguishable local environments. In c-Si/a-SiO$_2$ heterostructures, the same polarization layer captures the screened Hamiltonian response to an applied electric field, including its spatially nonuniform character at the atomic scale. Importantly, a physically motivated response mechanism alone does not guarantee quantitative screening: explicit field-response supervision substantially improves agreement with DFT and enables accurate band-edge predictions in the stronger-field test considered here. Together, these observations establish long-range electrostatics and external-field screening as connected, but separately testable, consequences of the polarization-aware Hamiltonian model.

Machine-learned Hamiltonians have already been integrated into device-scale transport calculations, as demonstrated by DeePTB--NEGF and HamGNN--NEGF~\citep{zou2025deeptbnegf,liu2026hamgnnnegf}. These studies establish the feasibility of large-scale quantum transport with learned Hamiltonians. Our focus is complementary: DeepH-Polar supplies an atomistically resolved, self-consistent description of built-in long-range electrostatics and field-induced screening, which is particularly relevant to devices containing heterogeneous interfaces, surfaces, and complex electrostatic environments. By combining Gaussian-orbital DFT Hamiltonian learning with environment-dependent atomic polarization, DeepH-Polar enables the electronic structure to respond to these fields rather than relying on a fixed zero-field Hamiltonian. Coupling the resulting Hamiltonian to an eigenstate-based free-carrier calculation enables electronic-structure simulations of an 8,194-atom gate-controlled device, including spatially resolved spectral features and interfacial charge distributions. This capability is particularly promising for studying interface roughness, structural defects, and other atomically heterogeneous features whose electronic effects are difficult to resolve in continuum descriptions.

\section{Methods}
\subsection{First-principles calculations and datasets}

Initial atomic configurations were constructed to sample crystalline, polar, surface, and disordered interfacial environments. The crystalline structures were obtained from the Materials Project~\citep{jain2013materialsproject} and expanded into supercells. Both GaN and AlN adopt the wurtzite structure. For GaN bulk, initial configurations were generated by varying the lattice constants within $\pm3\%$. GaN/AlN superlattices were constructed along the polar $z$ direction with different layer thicknesses and GaN-to-AlN ratios. GaN slabs were generated by varying the cleavage orientation, slab thickness, and vacuum spacing, with fractional-charge pseudo-hydrogen atoms used to passivate the exposed surfaces. For c-Si/a-SiO$_2$ heterostructures, amorphous-oxide interfaces were generated by a bond-switching Monte Carlo procedure and screened using valence-force-field (VFF) energies~\citep{wooten1985bond,zheng2025bsmc}. Both periodically continuous and hydrogen-terminated geometries along the interface-normal direction were included. The resulting structures contained 193--513 atoms.

The initial c-Si/a-SiO$_2$ structures were relaxed using the Orb-v3-direct-inf-omat machine-learning force field~\citep{rhodes2025orbv3}. The same force field was then used to perform 50\,ps of molecular dynamics for all initial configurations at 600\,K in the constant-particle-number, constant-volume, and constant-temperature ($NVT$) ensemble, using a Nos\'{e}--Hoover chain thermostat~\citep{martyna1992nosehoover}. Representative configurations were selected from the trajectories using the DImensionality-Reduced Encoded Clusters with stratified sampling (DIRECT) approach~\citep{qi2024direct,zhang2026mlipdata} for subsequent DFT calculations.

Reference Kohn--Sham Hamiltonian and overlap matrices for the datasets were calculated using the PWmat package accelerated by graphics processing units (GPUs)~\citep{jia2012gpu,jia2013multigpu}. The Perdew--Burke--Ernzerhof (PBE) exchange--correlation functional~\citep{perdew1996pbe} and Goedecker--Teter--Hutter (GTH) pseudopotentials parameterized for PBE~\citep{goedecker1996gth} were employed, with the DZVP-MOLOPT-SR-GTH Gaussian atomic-orbital basis~\citep{vandevondele2007gaussian}. The real-space integration cutoff was 862\,Ry, and Brillouin-zone sampling used a $k$-point spacing of approximately $0.06\pi\,\mathrm{\mathring{A}}^{-1}$ along the reciprocal lattice directions. The wavefunction and electron-density convergence tolerances were both $10^{-5}$. Spin--orbit coupling was not included. For field-dependent calculations, an external electric field was introduced explicitly through a prescribed external potential-energy profile along a direction containing a vacuum region.

Because Gaussian-type orbitals have no strictly finite spatial cutoff, the dataset Hamiltonian and overlap matrices were stored in atomic-pair blocks selected by an overlap-based threshold. For a pair of atoms $i$ and $j$ with $n_i$ and $n_j$ basis functions, respectively, we define the block root-mean-square overlap
\begin{equation}
\label{eq:method-overlap-cutoff}
t_{ij}=\left(\frac{1}{n_i n_j}\sum_{\mu=1}^{n_i}\sum_{\nu=1}^{n_j}\left|S_{i\mu,j\nu}\right|^2\right)^{1/2}.
\end{equation}
Only blocks with $t_{ij}>10^{-6}$ were retained, with the same atomic-pair selection applied to both $H$ and $S$.

\subsection{DeepH-Polar architecture and electrostatics}

DeepH-Polar employs a shared $E(3)$-equivariant DeepH-E3 backbone to predict both the short-range Hamiltonian and the atomic polarization parameters. Following the final message-passing layer, the updated edge features predict the short-range Hamiltonian $H_{\mathrm{SR},ij}$, while the updated node features predict the intrinsic atomic dipoles $\mathbf p_i^0$ from the $l=1$ components and the atomic polarizability tensors $\boldsymbol\alpha_i$ from the $l=0$ and $l=2$ components, corresponding to their isotropic and symmetric traceless parts, respectively. The long-range polarization layer couples these learned, environment-dependent quantities through electrostatics. For atomic positions $\{\mathbf R_i\}$ and dipole moments $\{\mathbf p_i\}$, we define a normalized Gaussian function and the corresponding polarization density:
\begin{align}
\label{eq:method-gaussian}
g_\sigma(\mathbf r)&=\frac{1}{(2\pi\sigma^2)^{3/2}}\exp\!\left(-\frac{|\mathbf r|^2}{2\sigma^2}\right), &
\mathbf P_\sigma(\mathbf r)&=\sum_j\mathbf p_j g_\sigma(\mathbf r-\mathbf R_j).
\end{align}
The polarization charge density $\rho_{\mathrm{pol}}=-\nabla\cdot\mathbf P_\sigma$ generates an electrostatic potential $\varphi_{\mathrm{pol}}$ satisfying Poisson's equation:
\begin{equation}
\label{eq:method-poisson}
-\varepsilon_0\nabla^2\varphi_{\mathrm{pol}}(\mathbf r)=\rho_{\mathrm{pol}}(\mathbf r),
\end{equation}
where $\varepsilon_0$ is the vacuum permittivity. The electrostatic potential and electric field are evaluated at the atomic positions; the Gaussian-regularized self-interaction is retained in the field kernel.

For the periodic boundary conditions used in training, solving Eq.~\eqref{eq:method-poisson} in reciprocal space gives the \emph{electron electrostatic potential energy} $\Phi_i^{\mathrm{per}}=-e\varphi_{\mathrm{pol}}(\mathbf R_i)$ and the dipole-interaction kernel $\mathbf K_\sigma$:
\begin{align}
\label{eq:method-periodic-phi}
\Phi_i^{\mathrm{per}}
&=\frac{e}{\varepsilon_0\Omega}\operatorname{Re}\sum_{\mathbf G\ne\mathbf0}
\frac{\mathrm i\,\exp(-\sigma^2G^2/2)}{G^2}
\sum_j(\mathbf G\cdot\mathbf p_j)
\exp[\mathrm i\mathbf G\cdot(\mathbf R_i-\mathbf R_j)],\\
\label{eq:method-periodic-k}
K_{\sigma,ij}^{\alpha\beta}
&=\frac{1}{\varepsilon_0\Omega}\sum_{\mathbf G\ne\mathbf0}
\frac{G_\alpha G_\beta}{G^2}\exp(-\sigma^2G^2/2)
\cos[\mathbf G\cdot(\mathbf R_i-\mathbf R_j)],
\end{align}
where $e>0$ is the elementary charge, $\Omega$ is the cell volume, and $\mathbf G$ denotes a reciprocal-lattice vector. The $\mathbf G=\mathbf0$ component is omitted to fix the mean electrostatic potential. The kernel is defined by the dipole-induced physical electric field $\mathbf E_i^{\mathrm{dip}}=-\sum_j\mathbf K_{\sigma,ij}\mathbf p_j$. Both quantities follow from the same Gaussian-broadened source and Poisson solution.

Beyond periodic conditions, the polarization layer also supports nonperiodic electrostatic boundary conditions, including Dirichlet and Neumann conditions. For a chosen boundary-condition set $B$, let $G_B(\mathbf r,\mathbf r')$ denote its electrostatic Green function, defined by $-\varepsilon_0\nabla_{\mathbf r}^2G_B(\mathbf r,\mathbf r')=\delta(\mathbf r-\mathbf r')$ with the corresponding boundary conditions. We define its source-smeared counterpart as
\begin{equation}
\label{eq:method-green-def}
\mathcal G_{B,\sigma}(\mathbf r,\mathbf R)
=\int_{\mathcal D}G_B(\mathbf r,\mathbf r')\,
 g_\sigma(\mathbf r'-\mathbf R)\,\mathrm d\mathbf r'.
\end{equation}
The electron potential energy and dipole-interaction kernel are then obtained from this same Green function:
\begin{align}
\label{eq:method-green-phi}
\Phi_{i,B}[\mathbf p;\sigma]
&=-e\sum_j\mathbf p_j\cdot\nabla_{\mathbf R_j}
\mathcal G_{B,\sigma}(\mathbf R_i,\mathbf R_j),\\
\label{eq:method-green-k}
K_{B,\sigma;ij}^{\alpha\beta}
&=\left.\frac{\partial^2\mathcal G_{B,\sigma}(\mathbf r,\mathbf R_j)}
{\partial r_\alpha\,\partial R_{j,\beta}}\right|_{\mathbf r=\mathbf R_i}.
\end{align}
For boundary conditions that break translational invariance, the kernel generally depends on the absolute positions of both source and observation points, rather than solely on their separation.

\subsection{Model training and loss functions}

We construct four datasets containing 598 GaN bulk, 349 GaN/AlN superlattice, 251 GaN slab, and 600 c-Si/a-SiO$_2$ structures; the latter also contains 591 pairs of DFT calculations under opposite applied fields. The datasets are divided into training, validation, and test subsets in an approximately 10:1:1 ratio, keeping configurations from the same molecular-dynamics trajectory within one subset whenever possible. We train nine models: DeepH-E3 and DeepH-Polar on each GaN dataset, and DeepH-E3, DeepH-Polar (H-only training), and field-supervised DeepH-Polar on c-Si/a-SiO$_2$. The first eight models use only zero-field Hamiltonian supervision; the last also learns the electric-field response. The H-only DeepH-Polar control is independently initialized and trained from scratch using only the zero-field structures in the c-Si/a-SiO$_2$ training subset, without finite-field response labels. The field-supervised batch comprises one zero-field structure together with zero, one, or two available positive--negative field pairs, whereas the other models use one structure per batch.

All models use a 7.2\,\text{\AA} neighbor cutoff, two hidden message-passing layers, $l_{\max}=4$ with parity enabled, an initial representation \texttt{64x0e}, and hidden representations \texttt{32x0e+16x1o+8x2e+}\allowbreak\texttt{4x3o+4x4e}. Training uses float32 precision and the Adam optimizer ($\beta_1=0.9$, $\beta_2=0.999$, $\epsilon=10^{-8}$) with an initial learning rate of $2\times10^{-3}$. A ReduceLROnPlateau scheduler acts on an exponentially smoothed validation loss (factor 0.5, patience 120, cooldown 40, threshold 0.05); training ends after 3000 epochs or when the learning rate reaches $3\times10^{-5}$.

The gauge-invariant losses in Eq.~\eqref{eq:loss} are implemented by removing an optimal uniform energy-reference shift for each supervised quantity $X$: the Hamiltonian ($X=H$) or its field derivative ($X=\Delta_\beta\equiv\partial H/\partial E_{\mathrm{ext},\beta}$). For the supervised matrix-element set $\mathcal M$, the shift and corresponding mean squared error are
\begin{align}
\label{eq:method-gi-loss}
c_X&=\frac{\sum_{(a,b)\in\mathcal M}S_{ab}\left(X^{\mathrm{pred}}_{ab}-X^{\mathrm{DFT}}_{ab}\right)}{\sum_{(a,b)\in\mathcal M}S_{ab}^2},\\
\mathcal L_X&=\frac{1}{|\mathcal M|}\sum_{(a,b)\in\mathcal M}\left(X^{\mathrm{pred}}_{ab}-X^{\mathrm{DFT}}_{ab}-c_X S_{ab}\right)^2.
\end{align}
For H-only training, $\mathcal L_{\mathrm{tot}}=\mathcal L_H$. In the field-supervised model, batches with field-response labels use
\begin{align}
\label{eq:method-mixed-loss}
\mathcal L_{\mathrm{tot}}&=0.7\mathcal L_H+0.3s_t\mathcal L_\Delta,\\
s_t&=0.95s_{t-1}+0.05r_t,\qquad
r_t\simeq\frac{\langle\mathcal L_H\rangle_{\mathrm{paired}}}{\langle\mathcal L_\Delta\rangle_{\mathrm{paired}}},
\end{align}
where $r_t$ is the ratio of the mean unweighted Hamiltonian loss to the mean unweighted field-response loss over samples with field pairs. The response loss is averaged over the available field directions for each structure. For batches without field pairs, the total loss reduces to $\mathcal L_{\mathrm{tot}}=\mathcal L_H$.

At fixed atomic geometry, the learned intrinsic dipoles and polarizabilities have no explicit applied-field dependence. Analytic differentiation of Eq.~\eqref{eq:polarization} gives, for a unit-field vector $\mathbf b_\beta$ in Cartesian direction $\beta$,
\begin{equation}
\label{eq:method-field-dipole}
(\mathbf I+\boldsymbol\alpha\mathbf K_\sigma)
\frac{\partial\mathbf p}{\partial E_{\mathrm{ext},\beta}}
=\boldsymbol\alpha\mathbf b_\beta.
\end{equation}
The Hamiltonian response $\partial H/\partial E_{\mathrm{ext},\beta}$ follows by analytic differentiation of Eqs.~\eqref{eq:potential}--\eqref{eq:hamiltonian-total}, rather than finite differences of predicted Hamiltonians. Its DFT supervision is constructed from the same atomic structure under opposite fields,
\begin{equation}
\label{eq:method-field-label}
\Delta_\beta^{\mathrm{DFT}}\simeq\frac{H^{\mathrm{DFT}}(+E_0\hat{\mathbf e}_\beta)-H^{\mathrm{DFT}}(-E_0\hat{\mathbf e}_\beta)}{2E_0},\qquad E_0=0.005\,\mathrm{V}/\text{\AA}.
\end{equation}
Each field pair provides one response target. All nine models were trained on two nodes equipped with eight NVIDIA RTX 3090 GPUs in total; the mean training times per epoch for the c-Si/a-SiO$_2$ DeepH-E3, H-only DeepH-Polar, and field-supervised DeepH-Polar models were 52.21, 61.29, and 99.27\,s, respectively.

\subsection{Self-consistent atomistic device calculations}

For device-scale calculations, the overlap matrix $S$ is evaluated directly from the Gaussian atomic orbitals rather than predicted by a learned model. Gaussian overlap integrals have analytic expressions and can be evaluated efficiently with optimized integral engines such as Libcint~\citep{sun2015libcint}; different atomic-pair blocks are readily parallelized. The device in Fig.~\ref{fig:device} uses donor concentrations of $10^{20}\,\mathrm{cm}^{-3}$ in the source and drain and an acceptor concentration of $10^{18}\,\mathrm{cm}^{-3}$ in the channel. The electrostatic problem imposes Dirichlet conditions at the gate and source--drain contacts, a zero-normal-field (Neumann) condition at the bottom insulating boundary, and periodic conditions in the lateral direction. The same boundary geometry is used both to solve the free-carrier/fixed-doping Poisson equation, Eq.~\eqref{eq:device-poisson}, and to evaluate the dipolar potential energy and interaction kernel through Eqs.~\eqref{eq:method-green-phi} and \eqref{eq:method-green-k}.

The chemical potential $\mu$ is determined by charge neutrality in the distant source--drain reservoirs $R$ of volume $V_R$,
\begin{equation}
\label{eq:device-mu}
2\sum_{i\in R}\!\left[\sum_{n\in C}w_{in}f(\varepsilon_n-\mu)-\sum_{n\in V}w_{in}\bigl(1-f(\varepsilon_n-\mu)\bigr)\right]=N_DV_R,
\end{equation}
where $N_D$ is the reservoir donor concentration, $C$ and $V$ denote the conduction- and valence-state manifolds, and $f(x)=[1+\exp(x/k_{\mathrm B}T)]^{-1}$. The corresponding spin-degenerate atomic populations are
\begin{align}
\label{eq:device-pop}
n_i^{\mathrm e}&=2\sum_{n\in C}w_{in}f(\varepsilon_n-\mu), &
n_i^{\mathrm h}&=2\sum_{n\in V}w_{in}\bigl[1-f(\varepsilon_n-\mu)\bigr].
\end{align}
These occupations yield $q_{i,\mathrm{free}}=e(n_i^{\mathrm h}-n_i^{\mathrm e})$, which includes only free electrons and holes relative to filled valence bands, not the valence bonding charge. The latter is already represented by the reference short-range Hamiltonian and the learned atomic polarization, so it is not added again to the device Poisson source.

For a fixed atomic geometry, $S$, $H_{\mathrm{SR}}$, and the $\Gamma$-point L\"owdin transformation $S^{-1/2}$ are computed and reused throughout the gate-voltage calculations. Within each Hamiltonian--charge self-consistent-field (SCF) iteration, FSM solves for electronic states within an energy window around the band edges by targeting the folded-spectrum operator,
\begin{equation}
\label{eq:device-fsm}
(\widetilde H-\omega\mathbf I)^2\mathbf v_n=(\varepsilon_n-\omega)^2\mathbf v_n.
\end{equation}
The working energy window extends from approximately 1\,eV below the average highest occupied molecular orbital (HOMO) energy of the preceding iteration to 1\,eV above its average lowest unoccupied molecular orbital (LUMO) energy and is divided into slices for iterative eigensolution. The resulting electron and hole occupations update the free-carrier charge, Poisson potential, and boundary-consistent atomic polarization, thereby closing the Hamiltonian--charge--potential cycle. For target states in each slice, the maximum residual of the original Hamiltonian eigenvalue equation must be below $10^{-6}\,$eV; the outer potential and atomic-charge residual tolerances are $10^{-5}\,$eV and $10^{-7}e$, respectively. Potential updates use linear mixing with direct inversion in the iterative subspace (DIIS) acceleration when applicable~\citep{pulay1982improvedscf}. Reported gate voltages include an approximately $+0.93\,$V flat-band correction $U_p-U_n$, determined from the chemical-potential difference between separately n- and p-doped bulk Si.

The 8,194-atom device contains 98,266 atomic orbitals, approximately $1.26\times10^6$ retained atomic-pair blocks and $1.88\times10^8$ matrix elements at the same overlap threshold $t_{ij}>10^{-6}$ used for the datasets. Overlap evaluation required 185\,s on 40 central processing unit (CPU) cores, and DeepH-Polar inference required 150\,s on 160 CPU cores. The one-time overlap eigendecomposition and L\"owdin matrix multiplications required 23.71 and 6.64\,min, respectively, using 960 CPU cores. At a representative reported gate bias of $0.73\,$V, the full SCF calculation converged in 20 iterations over approximately 4.14\,h using 512 CPU cores. CPU calculations used Intel Xeon Platinum 8375C processors.

\bibliographystyle{unsrtnat}
\bibliography{references}

\end{document}